\documentclass[traditabstract]{aa}
\usepackage{graphicx}
\usepackage{amsmath}
\usepackage{multirow}
\usepackage{natbib}
\usepackage{txfonts}
\usepackage{color}
\usepackage{footnote}
\usepackage{rotating}
\bibpunct{(}{)}{;}{a}{}{,}
\begin{document}

\renewcommand{\topfraction}{0.85}
\renewcommand{\bottomfraction}{0.7}
\renewcommand{\textfraction}{0.15}
\renewcommand{\floatpagefraction}{0.66}
\renewcommand{\vec}[1]{\mathbf{#1}}

\newcommand{\gammaray}{$\gamma$-ray}
\newcommand{\gammarays}{$\gamma$-rays}
\newcommand{\fluxunits}{cm$^{-2}$s$^{-1}$}

\def\gammavel{$\gamma^2$Velorum}
\def\gammav{$\gamma^2$Vel}
\def\sig{\hbox{$\sigma$}}
\def\flux{\textrm{TeV}^{-1}\textrm{cm}^{-2}\textrm{s}^{-1}}
\def\hess{H.E.S.S.}
\def\chandra{\emph{Chandra}}
\def\xmm{\emph{XMM-Newton}}
\def\asca{\emph{ASCA}}
\def\hone{H\,I}
\def\htwo{H\,II}

\def\psrb{PSR B1259-63/LS 2883}
\def\pulsar{PSR B1259-63}
\def\Hillas{\emph{Hillas}}
\def\model{\emph{model}}
\renewcommand{\deg}{\mbox{\ensuremath{^\circ}}}
\renewcommand{\arcmin}{\mbox{\ensuremath{^\prime}}}
\renewcommand{\arcsec}{\mbox{\ensuremath{^{\prime \prime}}}}
\newcommand{\beq}{\begin{equation}}
\newcommand{\enq}{\end{equation}}

\newcommand{\IS}[1]{#1}
\newcommand{\Ref}[1]{#1}
\newcommand{\Refnew}[1]{#1}

\def\vel{$\gamma^2$~Velorum}
\def\Ke{$K_\mathrm{e}$}

\title{Modeling of the Radio Emission from the Vela Supernova Remnant}

\author{I.~Sushch \inst{1,2,3}
\and B.~Hnatyk \inst{4}
}

\institute{
  Centre for Space Research, North-West University, Potchefstroom Campus, Potchefstroom, South Africa
\and
  Astronomical Observatory of Ivan Franko National University of Lviv, Lviv, Ukraine
\and
Humboldt Universit\"{a}t zu Berlin, Institut f\"{u}r Physik, Berlin, Germany
  \and
Astronomical Observatory of   Taras Shevchenko National University of Kyiv,  Kyiv, Ukraine
}

\date{Received 2 September, 2013; accepted 28 November, 2013}

\abstract
{Supernova remnants (SNRs) are widely considered to be sites of Galactic
cosmic ray (CR) acceleration. Vela is one of the nearest Galactic composite SNRs 
to Earth accompanied by the Vela pulsar and its pulsar wind nebula
(PWN) Vela X. The Vela SNR is one of the most studied remnants and it benefits from precise
estimates of various physical parameters such as distance and age. Therefore,
it is a perfect object for a detailed study of physical processes in SNRs.
The Vela SNR expands into the highly inhomogeneous cloudy interstellar
medium (ISM) and its dynamics is determined by the heating and evaporation
of ISM clouds. It features an asymmetrical X-ray morphology which is
explained by the expansion into two media with different densities.
This  could occur if the progenitor of the Vela SNR exploded close to
the edge of the stellar wind bubble of the nearby Wolf-Rayet star
$\gamma^2$Velorum and hence one part of the remnant expands into the
bubble. The interaction of the ejecta and the main shock of the remnant 
with ISM clouds causes formation
of secondary shocks at which additional particle acceleration takes place.
This may lead to the close to uniform distribution of
relativistic particles inside the remnant. We calculate the synchrotron radio
emission within the framework of the new hydrodynamical model which assumes the
supernova explosion at the edge of the stellar wind bubble. The simulated
radio emission agrees well with both the total radio flux from the remnant
and the complicated radio morphology of the source.} 


\authorrunning{I. Sushch \& B. Hnatyk}
\titlerunning{Modeling of the Radio Emission from Vela Supernova Remnant}
\maketitle

\section{Introduction}
\Ref{The} Vela Supernova remnant (SNR) is one of the most studied and closest SNRs to
the Earth. The \Ref{distance and the age} of the Vela SNR are \Refnew{determined well enough 
to make it a perfect object for the investigation of physical
processes}. Several estimates of the distance to the remnant exist (see \citet{vela_sushch}
and references therein), the most reliable of which is determined
from the VLBI parallax measurements of the Vela pulsar and is
$D_{\mathrm{Vela}} = 287^{+19}_{-17}$ pc \citep{dodson2003}. Equatorial
coordinates (J2000 epoch) of the Vela pulsar, which is assumed to be
situated in the geometrical center of the remnant, are $\alpha_{\mathrm{Vela}} = 08^h 35^m 20.66^s$
and $\delta_{\mathrm{Vela}} = -45\deg 10^\prime 35.2^{\prime\prime}$. 
The age of the Vela SNR is \Ref{usually determined as the 
characteristic age} of the Vela pulsar (PSR B0833-45) \Ref{which is about} $1.1\times10^{4}$ years \citep{reichley70}. 
\Ref{However, the characteristic age of the pulsar is estimated assuming 
that the pulsar spin-down braking index is equal to three (spin-down due 
to the magnetic dipole radiation) and that the initial rotational period 
is negligible in comparison to the current one (see e.g. \citet{gaensler&slane}). 
\citet{lyne_1996} estimated the braking index for the Vela pulsar to be very low of 
$1.4\pm0.2$, which may increase the estimate of the real age of the pulsar up to a 
factor of five comparing to the characteristic age. Meanwhile, an age estimate can 
be also obtained from the Vela SNR dynamics. A shock velocity $V_{\mathrm{sh}}$ of the middle-aged 
adiabatic SNR depends on the shock radius $R_{\mathrm{sh}}$ and the age $t_{\mathrm{age}}$ as $V_{\mathrm{sh}}=0.4R_{\mathrm{sh}}/t_{\mathrm{age}}$. In 
the case of the Vela SNR, we know both $R_{\mathrm{sh}}\sim 20$ pc (from the angular size and the 
distance to the Vela SNR) and $V_{\mathrm{sh}}\sim 660-1020$ km/sec (from the post-shock temperature 
of $0.5-1.2$ keV of the X-ray emitting gas) \citep{aschenbach95,vela_sushch}, which 
results in the hydrodynamical age of $t_{\mathrm{age}}=(0.7-1.2) \times 10^4$ y, which is close 
to the characteristic one.} 

Early radio observations of the Vela constellation
\citep{rishbeth_58} revealed three localized regions of enhanced brightness
temperature: Vela\,X, Vela\,Y and Vela\,Z. Vela\,X is the most intense
emission region which is believed to be a pulsar wind nebula (PWN) of
the Vela pulsar (see e.g. \citet{hess_velaX} and references
therein). It was first interpreted as a PWN associated with the Vela pulsar
by \citet{weiler&panagia_80}. Subsequent observations at 29.9, 34.5 and
408 MHz revealed one more region of intensified emission Vela\,W, which
features two peaks and is weaker than Vela\,Y and Vela\,Z \citep{alvarez2001}.
The spectral shape of the Vela\,W radio emission is similar to the spectral
shape of the radio emission from Vela\,Y and Vela\,Z suggesting the same nature of
these localized emission regions \citep{alvarez2001}.

The Vela SNR is one of the brightest sources on the X-ray sky.
The X-ray emission appears to be dimmer, but more
extended in the south-western (SW) part in comparison to the
north-eastern (NE) part of the remnant \citep{aschenbach95, lu&aschenbach_2000}.  
The bulk of the X-ray emission is distributed all over the SNR
without evidence of the main shock. Both features were recently
explained in \citet{vela_sushch} within the assumption that the
Vela SNR progenitor supernova exploded on the border of the stellar wind
bubble (SWB) of the nearby Wolf-Rayet (WR) star in the binary system \gammavel\
and that the remnant expands in a highly inhomogeneous, cloudy,
interstellar medium (ISM). Indeed, exploding at the border of
the SWB, the remnant  would expand into two media with different densities,
what, in turn, would cause a change of the X-ray luminosity and size
from the NE part to the SW part of the remnant. If the remnant expands
into the cloudy ISM with the high ratio of the clouds' volume averaged number density 
to the intercloud number density, its dynamics and X-ray emission would be determined 
mostly by the matter initially concentrated in clouds \citep{white&long91}. Due to a two-component 
core-corona structure of clouds in the Vela SNR \citep{miceli_2006}, the heating and evaporation of 
clouds results in the two-component structure of the remnant's interior. 
The hot evaporated gas component with the volume filling factor
close to unity dominates the shock dynamics, while the cooler and denser
component with the filling factor close to zero dominates
the X-ray radiation from the remnant. The role of the initial intercloud ISM gas is
negligible \citep{vela_sushch}.

\gammavel\ is the WC8+O8-8.5III binary system whose WR component (WR11) is the closest
WR star to the Earth. There are several recent estimates of the distance
to \gammavel\ which are based on different measurements, but reveal similar results.
\citet{millour2007} provide an interferometric estimate of the distance of
$D_{\gamma^2\mathrm{Vel}} = 368^{+38}_{-13}$ pc. Based on the orbital solution for the \gammavel\ binary
obtained from the interferometric data \citet{north2007} calculated the distance at
$D_{\gamma^2\mathrm{Vel}} = 336^{+8}_{-7}$ pc. Finally, the estimate of the distance based on HIPPARCOS parallax
measurements is $D_{\gamma^2\mathrm{Vel}} = 334^{+40}_{-32}$ pc \citep{van_leeuwen2007}. The latter is used for
calculations in this paper. Equatorial coordinates (J2000 epoch) of \gammavel\ are
$\alpha_{\gamma^2\mathrm{Vel}} = 08^h 09^m 31.95^s$ and
$\delta_{\gamma^2\mathrm{Vel}} = -45\deg 20^\prime 11.7^{\prime\prime}$ \citep{van_leeuwen2007}.

In this paper, we present a simulation of the radio synchrotron emission
from the Vela SNR. The simulation was performed in the framework of the
\citet{vela_sushch} model using estimates of physical parameters of
the Vela SNR and its interior derived in that work (see
Table\,\ref{phys_params}). The nucleon number densities $n$, corresponding
filling factors $f$ and the kinetic gas temperatures $T$ are presented
for both hot and cool components of the remnant's interior.
The simulated radio emission from the remnant is
compared to the observational data presented in \citet{alvarez2001}.

The paper is
structured as follows: in Section \ref{geom}, the geometrical model
of the Vela SNR$-$\vel\ system is presented. In Section
\ref{radio}, the synchrotron radio emission from the spherical SNR
with the uniform distribution of electrons is investigated, which is
applied then to the Vela SNR in Section \ref{radio_Vela} assuming the
Vela SNR as a combination of two hemispheres with different radii.
The morphology and the overall flux of the radio emission are discussed
and compared to observational data. Finally, results are discussed in
Section \ref{discussion} and summarised in Section \ref{sum}.

\begin{table}
\centering
\caption{Physical parameters of the Vela SNR derived in \citet{vela_sushch}}
\begin{tabular}{ p{5cm} | c c}
\hline
\hline
Parameter & NE & SW \\
\hline
Explosion energy $E_{\mathrm{SN}}$  [erg]&  \multicolumn{2}{c}{$1.4\times10^{50}$ } \\
Radius $R_{\mathrm{Vela}}$ [pc]& 18& 23 \\
\hline
Hot component: && \\
$n_{\mathrm{hot}}$ [cm$^{-3}$] & 0.04 & 0.01 \\
$f_{\mathrm{hot}}$ & 0.93&  0.91\\
$T_{\mathrm{hot}}$ [K] & $9 \times 10^6$ & $1.5 \times 10^7$ \\
\hline
Cool component: && \\
$n_{\mathrm{cool}}$ [cm$^{-3}$] & 0.38 & 0.10 \\
$f_{\mathrm{cool}}$ & 0.07& 0.09 \\
$T_{\mathrm{cool}}$ [K] & $1 \times 10^6$ & $1.7 \times 10^6$ \\
\hline
\end{tabular}
\label{phys_params}
\end{table}

\section{Geometrical Model}
\label{geom}
As shown in \citet{vela_sushch}, if the radius of the stellar wind bubble (SWB) around
\vel\ is about 30-70 pc it should physically intersect with
the Vela SNR \Refnew{which} would cause the change of physical properties
of the remnant in the part which expands into the SWB. It was
suggested that the progenitor supernova exploded on the border of
the SWB \Refnew{which} naturally explained the step-like change in properties
from the NE to the SW part of the remnant. Expanding into media 
with different densities, the Vela SNR can be aproximated as a combination 
of two hemispheres, south-western (SW) and north-eastern (NE), with radii 
$R_{\mathrm{SW}} = 23$\,pc and $R_{\mathrm{NE}} = 18$\,pc, respectively 
\citep{vela_sushch}. However, in order to explain the complicated morphology 
of the source a detailed geometrical model of the system is required.

\begin{figure}
  \centering
  \resizebox{\hsize}{!}{\includegraphics[width=\textwidth]{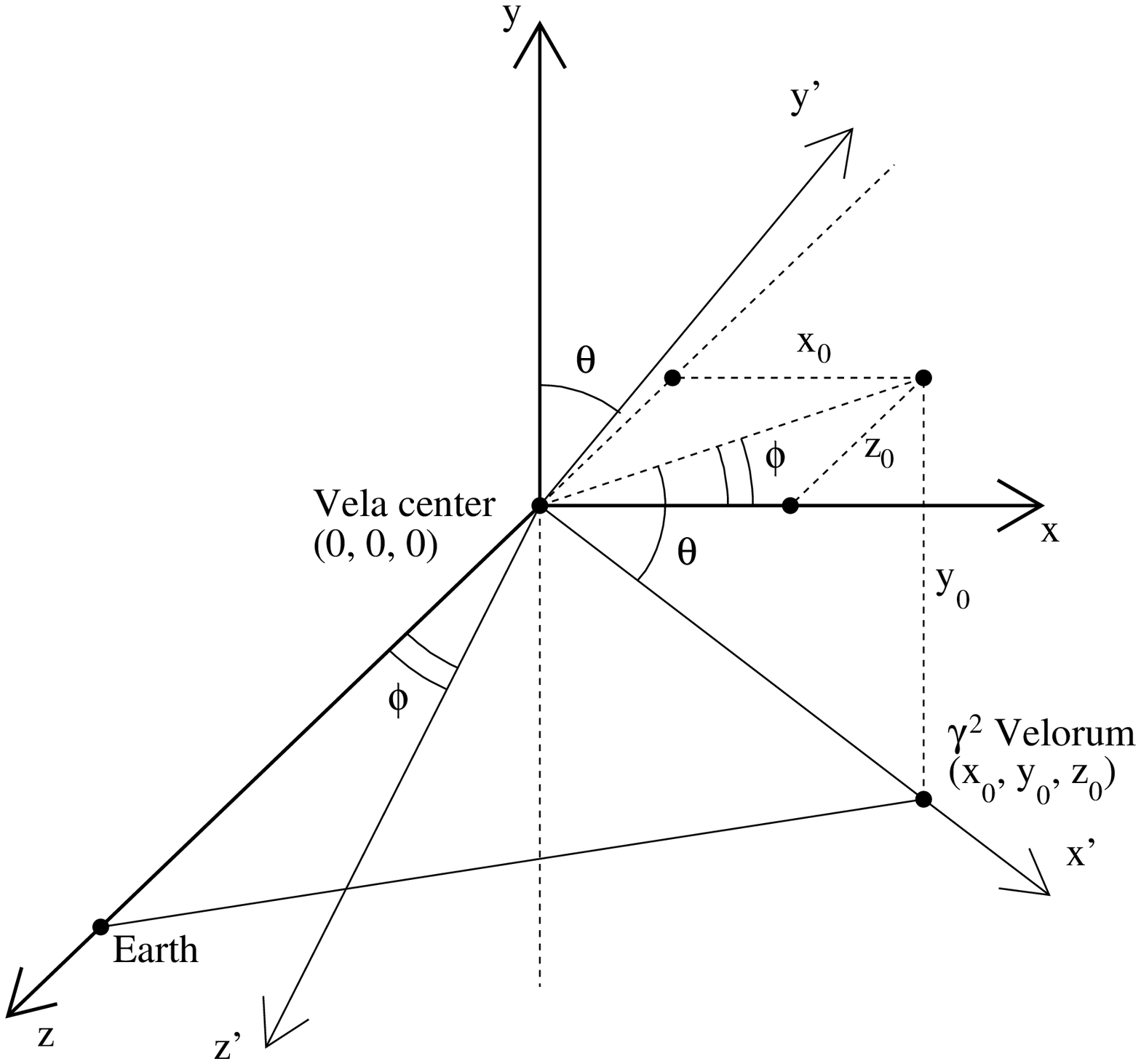}}
  \caption{Definition of coordinate systems $\vec{K}$ and $\vec{K^{\prime}}$. See text for explanation.}
  \label{geom_model}
\end{figure}

Let us define a coordinate frame $\vec{K}$ by its origin at the center of the
Vela SNR, the $z$-axis coinciding with the direction towards Earth, the
$y$-axis tangent to a line of declination and the $x$-axis tangent to a
circle of right ascension of the celestial sphere with the radius
$r = D_{\mathrm{Vela}}$ (Fig. \ref{geom_model}). The $xy$-projection of the Vela SNR can be then easily
converted into equatorial coordinates using coordinate
transformations
\begin{align}
\label{eq_coord_trans}
x &= D_{\mathrm{Vela}} \cos\delta_{\mathrm{Vela}} \sin(\alpha_{\mathrm{Vela}} - \alpha), \nonumber \\
y &= D_{\mathrm{Vela}} \sin(\delta - \delta_{\mathrm{Vela}}),
\end{align}
assuming that $(\alpha-\alpha_{\mathrm{Vela}})$ and
$(\delta - \delta_{\mathrm{Vela}})$ are small.

\begin{figure*}
  \centering
  \includegraphics[width=\linewidth]{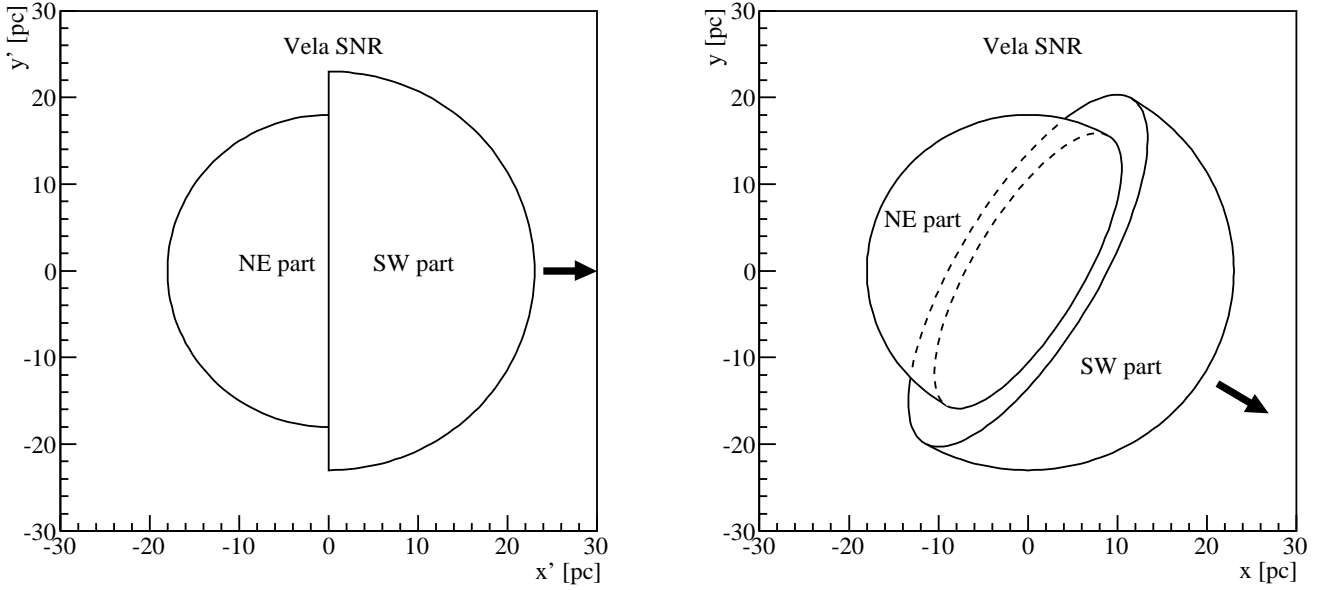}
  \caption{A schematic illustration of the $xy$-projection of the Vela SNR in $\vec{K^\prime}$ (left) and $\vec{K}$ (right) coordinate systems. The
direction towards the position of \gammavel\ is shown with an arrow.}
  \label{vela_geom_cartoon}
\end{figure*}

A $\vec{K^{\prime}}$ frame is defined, in turn, by its origin in the 
center of the Vela SNR, $x^{\prime}$-axis coinciding with the direction towards 
\gammavel\ and $y^{\prime}$ and $z^{\prime}$ axes defined in a way that the 
$y^{\prime}z^{\prime}$-plane separates NE and SW hemispheres of the Vela SNR 
(Fig. \ref{vela_geom_cartoon} left panel). The $\vec{K^{\prime}}$ frame 
can be transformed to the $\vec{K}$ frame by the rotation as 
\beq
\label{eq_rot}
\vec{K^\prime} = R_{z^\prime}(\theta)R_{y}(\phi)\vec{K} = \begin{bmatrix}
            \cos\theta \cos\phi  & \sin\theta & \cos\theta \sin\phi \\
            -\sin\theta \cos\phi & \cos\theta & -\sin\theta \sin\phi \\
            -\sin\phi            & 0          & \cos\phi
          \end{bmatrix}
\vec{K},
\enq
where $R_{z^\prime}(\theta)$ and $R_{y}(\phi)$ are rotation matrices for the
rotation around $z^\prime$-axis through angle $\theta$ and around 
$y$-axis through angle $\phi$ respectively. In Fig.\,\ref{vela_geom_cartoon}, 
projections of the Vela SNR on the xy-plane in $\vec{K^{\prime}}$ and $\vec{K}$ 
coordinate systems are shown schematically. The xy-projection in the $\vec{K}$ coordinate 
system reflects how the remnant is seen on the sky by the observer. 
It can be transformed into the equatorial coordinate system using 
coordinate transformation equations (Eqs.\,\ref{eq_coord_trans}).

For given coordinates $(x_0, y_0, z_0)$ of \gammavel\ in the $\vec{K}$-frame
the rotation angles $\phi$ and $\theta$ can be estimated as
\begin{align}
\label{eq_angles}
\tan\phi &= \frac{\lvert z_0 \rvert}{\lvert x_0 \rvert}, \nonumber \\
\tan\theta &= \frac{\lvert y_0 \rvert}{\sqrt{x_0^2 + z_0^2}}.
\end{align}

\begin{figure*}
  \centering
  \includegraphics[width=\linewidth]{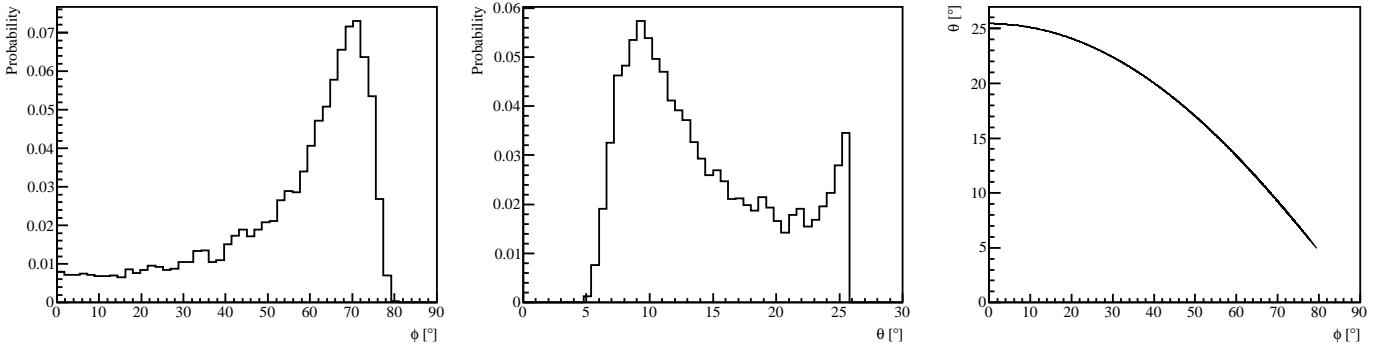}
  \caption{Distributions of rotation angles $\phi$ (left panel) and $\theta$ (central panel) obtained for the known estimates of distances
to the Vela SNR and \gammavel\ (see text for explanation). On the right panel the mutual dependence of rotation angles is shown.}
  \label{rotation_angles}
\end{figure*}
In turn, $x_0$, $y_0$ and $z_0$ can be calculated using known distances and
equatorial coordinates of the sources by transformation equations: 
\begin{align}
\label{eq_coord_trans_2}
x_0 &= D_{\gamma^2\mathrm{Vel}} \cos\delta_{\mathrm{Vela}} \sin(\alpha_{\mathrm{Vela}} - \alpha_{\gamma^2\mathrm{Vel}}), \nonumber \\
y_0 &= D_{\gamma^2\mathrm{Vel}} \sin(\delta_{\gamma^2\mathrm{Vel}} - \delta_{\mathrm{Vela}}), \\
z_0 & = D_{\mathrm{Vela}} - D_{\gamma^2\mathrm{Vel}}\cos\Delta, \nonumber
\end{align}
where $\Delta$ is the angular distance between Vela and \gammavel\ which is given by
\beq
\cos\Delta = \sin\delta_{\mathrm{Vela}} \sin\delta_{\gamma^2\mathrm{Vel}} + \cos\delta_{\mathrm{Vela}} \cos\delta_{\gamma^2\mathrm{Vel}} \cos(\alpha_{\mathrm{Vela}} - \alpha_{\gamma^2\mathrm{Vel}}).
\enq
Assuming that estimates of distances to the Vela SNR ($D_{\mathrm{Vela}} = 287^{+19}_{-17}$ pc) and
\gammavel\ ($D_{\gamma^2\mathrm{Vel}} = 334^{+40}_{-32}$ pc) follow asymmetric Gaussian distributions
and asymmetric errors correspond to standard deviations of the distribution, one can obtain
distributions of the rotation angles $\phi$ and $\theta$ from Eqs. (\ref{eq_angles}-\ref{eq_coord_trans_2})
by varying distance estimates. Angle distributions presented in Fig. \ref{rotation_angles} (left and middle panels)
show the probability of the true angle $\phi_{\mathrm{true}}/\theta_{\mathrm{true}}$ being in the range of angles
$(\phi+\delta\phi)/(\theta+\delta\theta)$. Each histogram contains $50$ bins, i.e. $\delta\phi = 1.8^{\circ}$
and $\delta\theta = 0.6^{\circ}$. By calculating the mode for each distribution the most
probable values of rotation angles $\phi$ and $\theta$ can be estimated
\begin{align}
\phi_0 & = 71.1\deg \pm 0.9\deg, & \theta_0 & = 9.3\deg \pm 0.3\deg.
\end{align}
The angles $\phi$ and $\theta$ are correlated and their mutual dependence
is shown in the right panel of Fig. \ref{rotation_angles}.

\section{Radio emission from the spherical SNR with the uniform electron distribution}
\label{radio}


\Ref{The radio emission from the Vela SNR shows an indication of the 
brightening towards the center which is not usually expected in the 
shell-like SNRs, where electrons emitting synchrotron radiation are 
accelerated at the main shock and are concentrated close to the edge of the remnant.}
\Ref{In the case of the Vela SNR,} the radio luminosity
grows towards the center of the remnant featuring several localized maxima
within the SNR. Such morphology suggests a close to uniform distribution
of relativistic electrons inside the remnant. Possible reasons for such
a distribution of electrons are discussed in the Section \ref{discussion}. 
In this section and the following one we investigate the radio emission
from the SNR with a uniform distribution of relativistic electrons and apply
this model to the case of the Vela SNR, considering it as a composition of two
hemispheres with uniform distribution of relativistic electrons and magnetic
field in each of them.

We assume that the distribution of the relativistic electron density 
$N_\mathrm{e}(\gamma)$ with energies follows a power law
\beq
\frac{dN_\mathrm{e}}{d\gamma} = K_\mathrm{e} \gamma^{-p}, \gamma\geq\gamma_{\mathrm{min}},
\enq
where $\gamma$ is the electron Lorentz factor, $\gamma_{\mathrm{min}}$ is the minimal
electron Lorentz factor, $K_\mathrm{e}$ is the normalization constant and $p$ is the electron spectral index.
Then the overall synchrotron flux density at frequency $\nu$ from
the spherical SNR located at distance $D$ can be calculated
as \citep{rybicki&lightman}
\begin{align}
\label{eq_Snu}
& S_\nu = \nonumber \\
&\frac{R^{3}}{3D^{2}} K_\mathrm{e}
\frac{\sqrt{3}q^{3}B \sin{\theta}}{mc^{2}(p+1)}
\Gamma \left(\frac{p}{4}+\frac{19}{12}\right)
\Gamma \left(\frac{p}{4}-\frac{1}{12}\right)
\left(\frac{2\pi mc}{qB\sin{\alpha}}\nu\right)^{-\frac{(p-1)}{2}},
\end{align}
where $B$ is the magnetic field, $R$ is the radius of the SNR, $q$ is the
electron charge, $m$ is the electron mass, $c$ is the speed of light and
$\alpha$ is the angle between the magnetic field and electron velocity.
It is assumed that electron velocities are isotropically distributed and
a root mean square value $\sin{\alpha}=\sqrt{2/3}$ can be used.

The flux density depends on three parameters, namely radius of the SNR $R$,
magnetic field inside the SNR $B$ and constant $K_\mathrm{e}$. If the distance
to the remnant is known the radius can be calculated directly from
the angular size of the SNR.

The interior magnetic field is assumed to be determined mainly by shock-cloud
interactions which result in vorticity and turbulence generation (see numerical
calculations of shock-cloud interactions in \citet{inoue2011} and references therein).
It was shown in \citet{inoue2011} that magnetic field amplification is determined
by saturation at $\beta=8\pi P_{\mathrm{gas}}/B^2\sim 1$
(the equilibrium condition of the magnetic pressure and the thermal pressure
of particles $P_{\mathrm{gas}}$).
In the case of the Vela SNR the evaporated cloud material with spatially nearly uniform
thermal pressure fills up practically all the volume of the remnant, therefore the
magnetic field will be uniform within the remnant and can be estimated as
\beq
\label{eq_B}
B=\sqrt{8 \pi n_{\mathrm{tot}} k_\mathrm{B} T},
\enq
where $n_{\mathrm{tot}}=n/\mu$ is the total number density of electrons and nuclei, $n$ is the nucleon number density, $\mu=16/27$ is the molar mass, 
and $T$ is the kinetic gas temperature inside the remnant.

Finally, to calculate \Ke\ one should know the total energy in relativistic electrons
$E_{\mathrm{e}}$ and the size of the remnant. The total energy in electrons is given by the
integration of the electron energy spectrum over all electron energies and over the
volume of the remnant

\beq
E_{\mathrm{e}}=\iint mc^{2}\gamma\frac{dN_{\mathrm{e}}}{d\gamma} d\gamma dV =
\frac{4}{3}\pi R^{3}mc^{2}K_{\mathrm{e}} \int_{\gamma_{\mathrm{min}}}^\infty \gamma^{-p+1} d\gamma.
\enq
Then parameter $K_\mathrm{e}$ can be expressed as
\beq
\label{eq_Ke}
K_{\mathrm{e}} = \frac{E_\mathrm{e}}{\frac{4}{3}\pi R^{3} m c^{2} \int_{\gamma_{\mathrm{min}}}^\infty \gamma^{-p+1} d\gamma},\,(p>2).
\enq

\section{Radio emission from the Vela SNR}
\label{radio_Vela}

We assume that the explosion of the supernova was spherically symmetrical.
In this scenario the energy transferred to relativistic electrons in the SW and
NE parts of the remnant would be the same and equal to a half of the total
energy in electrons $E_{\mathrm{e}}$. Since the SW and NE parts of the remnant
have different size, relativistic electron densities in these parts would be also different

\begin{align}
N_{\mathrm{e,\,SW}}(\gamma \ge \gamma_{\mathrm{min}}) & = K_{\mathrm{e,\,SW}} \int_{\gamma_{\mathrm{min}}}^\infty \gamma^{-p} d\gamma, \nonumber \\
N_{\mathrm{e,\,NE}}(\gamma \ge \gamma_{\mathrm{min}})  & = K_{\mathrm{e,\,NE}} \int_{\gamma_{\mathrm{min}}}^\infty \gamma^{-p} d\gamma,
\end{align}
where parameter $K_{\mathrm{e,\,SW/NE}}$ is dependent on the size of the hemisphere and can be estimated
from Eq. \ref{eq_Ke} for $R_{\mathrm{SW/NE}}$. We assume that the minimal energy of
electrons is $\gamma_{\mathrm{min}}mc^2 = 100$ MeV. The total energy in electrons $E_{\mathrm{e}}$ and
the electron spectral index can be derived from the observational data as discussed below.
Magnetic fields inside the remnant $B_{\mathrm{NE/SW}}$ can be calculated using
Eq.\,(\ref{eq_B}) and estimates of nucleon number density $n_{\mathrm{hot}}^{\mathrm{NE/SW}}$ and kinetic
temperature $T_{\mathrm{hot}}^{\mathrm{NE/SW}}$ of the hot component (dominant across the remnant) listed
in the Table \ref{phys_params}.
In the NE part of the remnant the magnetic field
is $B_{\mathrm{NE}} = 46\,\mu$G, while in the SW part it is \Ref{$B_{\mathrm{SW}} = 30\,\mu$G}.


\subsection{Integrated flux density}

\begin{figure}
  \centering
  \includegraphics[width=\linewidth]{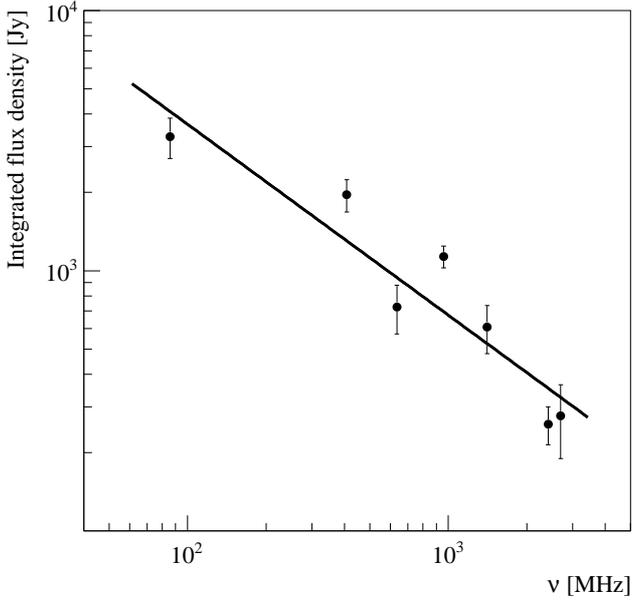}
  \caption{Sum of the integrated flux density spectra from Vela Y and Vela Z,
presented in \citet{alvarez2001}. The straight line represents a model fit to the data.}
  \label{flux_density_VelaYZ}
\end{figure}

By fitting the model flux density (Eq. (\ref{eq_Snu})) to the observational data
one can calculate the total energy in electrons $E_{\mathrm{e}}$ and the electron
spectral index $p$ for the assumed minimal electron Lorentz factor
$\gamma_{\mathrm{min}}$.
\citet{alvarez2001}  provide the flux density from the whole remnant $S_{XYZ}$ and
flux densities from localized emission regions $S_{X}$, $S_{Y}$ and $S_{Z}$  from Vela\,X,
Vela\,Y and Vela\,Z, respectively (see Tab. 2 therein). The ratio of the integrated flux density
$S_{XYZ}$ to the  sum of components $S_{X}+S_{Y}+S_{Z}$  shows appropriate self-consistency.
Vela\,X is the PWN associated with the Vela pulsar and should not be taken into account
for the study of the emission from the Vela SNR itself. Therefore, the flux density from the
whole remnant $S_{XYZ}$ which includes the emission from the PWN Vela\,X cannot be used here. 
The emission from Vela\,Y and Vela\,Z comes mainly from the NE part of the remnant.
Flux densities of Vela\,Y and Vela\,Z were summed up and the resulting cumulative flux
density (Fig. \ref{flux_density_VelaYZ}) is assumed to be the flux density of the NE part
of the Vela SNR. 
The model fit of the data (solid line in Fig. \ref{flux_density_VelaYZ})
results in 
the following values for the fitting parameters\footnote{The
distance to the remnant was fixed in these calculations and errors of the distance estimate were not taken
into account. Therefore estimated errors on the parameters might be underestimated.}
\begin{align}
  p &= 2.47 \pm 0.09, \\
  E_{\mathrm{e}} &= (2.4 \pm 0.2) \times 10^{47} \, \mathrm{erg},
\end{align}
which corresponds to a fraction of the total explosion energy transferred to
electrons of
\beq
\epsilon_{\mathrm{e}} = E_{\mathrm{e}}/E_{\mathrm{SN}} = (1.7 \pm 0.1) \times 10^{-3}
\enq
which is close to a typical value, expected
for SNRs (see e.g. \citet{katz_waxman_2008, gabici_snr_review}). 
Flux densities at 29.9 MHz and 34.5 MHz were not fit since
they show some indication of absorption \citep{alvarez2001}. For the known $p$ and $E_{\mathrm{e}}$, the relativistic electron
densities and parameters $K_{\mathrm{e}}$ for both, NE and SW, parts of the remnant can be calculated. They are presented in
Table \ref{electron_params}.

\begin{table}
\centering
\caption{Physical parameters of the relativistic electron population inside the Vela SNR.}
\begin{tabular}{ p{3.5cm} | c | c c}
\hline
\hline
Obtained from & Parameter & NE & SW \\
\hline
\multirow{5}{*}{a fit of the radio spectrum}& $p$  &  \multicolumn{2}{c}{$2.47\pm0.09$} \\
                                              & $E_{\mathrm{e}}$\,[erg] & \multicolumn{2}{c}{$(2.4\pm0.2)\times10^{47}$} \\
                                              & $\epsilon_{\mathrm{e}}$ & \multicolumn{2}{c}{$(1.7\pm0.1)\times10^{-3}$}  \\
                                              & $K_{\mathrm{e}}$\,[cm$^{-3}$] & $2.4\times10^{-6}$& $1.2\times10^{-6}$\\
                                              & $N_{\mathrm{e}}$\,[cm$^{-3}$] & $0.7\times10^{-9}$& $0.3\times10^{-9}$\\
\hline
\multirow{4}{*}{a flux density at 408 MHz}& $E_{\mathrm{e}}$\,[erg] & \multicolumn{2}{c}{$(3.6\pm0.5)\times10^{47}$} \\
                                              & $\epsilon_{\mathrm{e}}$ & \multicolumn{2}{c}{$(2.6\pm0.3)\times10^{-3}$}  \\
                                              & $K_{\mathrm{e}}$\,[cm$^{-3}$] &  $3.6\times10^{-6}$&  $1.7\times10^{-6}$\\
                                              & $N  _{\mathrm{e}}$\,[cm$^{-3}$] & $1.1\times10^{-9}$ & $0.5\times10^{-9}$\\
\hline
\end{tabular}
\label{electron_params}
\end{table}

\subsection{Morphology}

The brightness temperature map of the Vela SNR at 408 MHz in equatorial coordinates was simulated using the geometrical model
presented in Section \ref{geom}. The emission was modelled in 3D in the $\vec{K^{\prime}}$ coordinate system, treating every unit
volume as a separate emitter. Then the 3D-model of the remnant was converted to the $\vec{K}$ coordinate system and projected onto
the $xy$-plane. Finally, the $xy$-plane was converted to the equatorial coordinates as described in the Section \ref{geom}.

\begin{figure*}
  \centering
  \includegraphics[width=\linewidth]{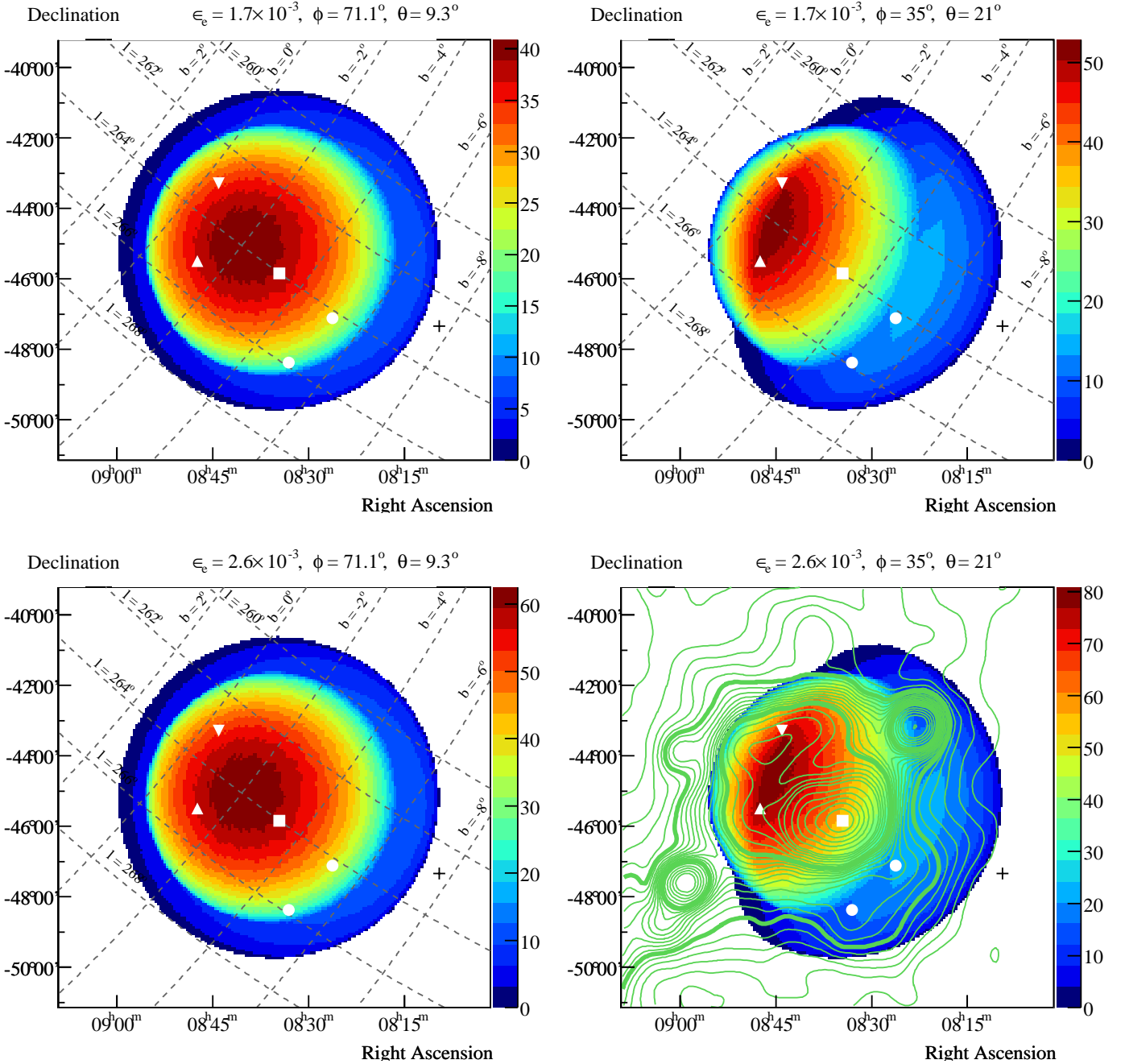}
  \caption{Simulated brightness temperature maps at 408 MHz \Ref{in equatorial coordinates overlaid by galactic coordinates}. Maps are calculated for $\epsilon_{\mathrm{e}} = 1.7\times10^{-3}$ (two upper panels) 
and $\epsilon_{\mathrm{e}} = 2.6\times10^{-3}$ (two lower panels). Two sets of rotation angles are considered: most probable values
$\phi = 71.1\deg$, $\theta = 9.3\deg$ (two left panels) and values which describe observational data the best $\phi = 35\deg$,
$\theta = 21\deg$ (two right panels). The color reflects the brightness temperature in K. \Ref{The angular resoulution of the modeled brightness temperature 
distributions is $4^{\prime}$.} In each map
\Ref{white} filled down- and up-pointing triangles denote peak locations of Vela\,Y an Vela\,Z correspondingly, two \Ref{white} filled circles show locations of
two peaks of Vela\,\Ref{W}, a \Ref{white} filled square denotes the position of the Vela\,X peak \Ref{\citep{alvarez2001}}, and a cross reflects the location of \gammavel. 
\Ref{The bottom right map which represents "the best fit" scenario is overlaid with observed 408 MHz radio contours with $51^{\prime}$ angular resolution from \citet{408_atlas}. 
The contours represent the brightness temperature in K and the steps are 4\,K from 40\,K to 100\,K, 10\,K from 100\,K to 150\,K and 25\,K further on, the contours of 60 and 100\,K 
are shown with the bold lines. Contrary to the map presented in \citet{alvarez2001}, the galactic background is not removed, which explains the difference in values of the brightness temperature. 
\citet{alvarez2001} adopted two background temperatures of 50 and 60\,K at 408 MHz. Besides Vela X in the centre, additional prominent (RCW 38 ($\alpha=08^h59^m, \delta = -47^{\circ}32^{\prime}$), 
Puppis A ($\alpha=08^h23^m, \delta=-42^{\circ}42^{\prime}$)) and weaker compact sources (RCW 36, RCW 33 and RCW 27 clockwise along the NE-North surface) are visible.}}
  \label{bright_temp}
\end{figure*}

Primarily, the simulation was performed using estimates of the physical parameters of the electron population obtained from the fit
of the observed radio spectrum (see the subsection above) and the mode values $\phi_0$ and $\theta_0$ of rotation angles $\phi$ and $\theta$. In the top
left panel of Fig. \ref{bright_temp} the simulated temperature brightness map for these values is shown. The color corresponds
to the brightness temperature in K. The brightness temperature distribution
is determined by the integration of the radio emission along the line of sight.
The radio emission is stronger in the NE hemisphere of the remnant due to the higher
density of relativistic electrons and stronger magnetic field. The emission would peak
in the \Ref{center} of the remnant
in the two limiting cases, namely, when the \Ref{center} of the Vela SNR and \gammavel\ are
located on the same line of sight, i.e. $\phi = 90\deg$ and $\theta = 0\deg$
and in the case when the Vela SNR center$-$\gammavel\ symmetry axis is
perpendicular to the line of sight, i.e. the Vela SNR and \gammavel\
are located at the same distance and $\phi = 0\deg$.
In the intermediate case $0<\phi<90\deg$ the peak of the
brightness temperature is shifted to the NE part of the remnant,
as shown in Fig. \ref{bright_temp} (top left panel) for the
case of the most probable values of $\phi = 71.1\deg$ and $\theta=9.3\deg$.


For $\phi \leq 40\deg$, a second, considerably fainter, peak appears in the SW part of the remnant. It is
not seen for $\phi>40\deg$ due to the overlapping effect, which causes the contamination of the SW part of the
remnant by the radio emission from the NE hemisphere. Remarkably, these two theoretically predicted peaks
correspond to the observed morphology of the brightness temperature distribution in the
Vela SNR - the existence of "hot spots" in both, the NE (two close localized regions of Vela\,Y and Vela\,Z)
and SW (two peaks of Vela\,W) parts of the remnant. The peaks of the emission regions Vela\,Y and Vela\,Z
are shown as down- and up-pointing triangles, respectively, and the peaks of the Vela\,W region are shown
as filled circles in each map in Fig. \ref{bright_temp}. For another combination of angles
$\phi = 35\deg$ and $\theta = 21\deg$, which is also compatible with estimates for the
distances to the Vela SNR and \gammavel, the positions of the simulated brightness temperature
peaks coincide with the observed localized regions (Fig \ref{bright_temp} top right), suggesting
that the complicated morphology of the Vela SNR might be a result of superimposed emission
in the system with a specific spatial orientation.
Modeled peak brightness temperatures on the top right panel of Fig. \ref{bright_temp} are slightly
lower than the observed brightness temperatures of the Vela\,Y, Vela\,Z,  and Vela\,W peaks. As reported by
\citet{alvarez2001} the brightness temperature of the Vela\,Y and Vela\,Z peaks is about 90\,K and the brightness
temperature of Vela\,W peaks is 35$-$40 K,while on the simulated map peak temperatures are
$\sim50 $ K and $\sim15$ K, respectively. This difference is expected given that the observed cumulative
Vela\,Y and Vela\,Z flux density at 408 MHz is $1.5$ times higer than the model fit to the data at this frequency
(Fig. \ref{flux_density_VelaYZ}).
Therefore, in order to be able to accurately compare the simulated and observed brightness
temperature distributions one has to derive the total energy in electrons directly from
the observed flux density at 408\,MHz. The spectral index is assumed to be $p = 2.47$ as
obtained in a fit. The derived physical parameters of the electron population are presented
in Table \ref{electron_params}. Using these new estimates we simulate the brightness
temperature distribution for the two sets of $\phi$ and $\theta$ discussed above
(Fig. \ref{bright_temp} bottom left and bottom right panels). In this case, the simulated
peak brightness temperatures for the combination of angles $\phi = 35\deg$ and $\theta = 21\deg$
(Fig. \ref{bright_temp} bottom right) are in a good agreement with the observational results.
The brightness temperature of the NE peak is about 80 K and the brightness temperature of
the SW peak is about 25 K. This consistency with the observational data is another
confirmation of the validity of our model.

\section{Discussion}
\label{discussion}

\subsection{Uniform distribution of relativistic electrons}
A typical middle-aged SNR in the adiabatic stage of evolution is a powerful source of
both thermal X-ray and nonthermal synchrotron radio emission. A strong shock wave
compresses and heats the interstellar gas up to keV temperatures creating a shell-like
X-ray morphology due to the concentration of the shocked plasma downstream of the shock front.
At the same time the shock wave accelerates charged particles, electrons, protons and nuclei,
to ultrarelativistic energies via the diffusive shock acceleration mechanism. Since both the
magnetic field and relativistic electrons are also concentrated downstream the shock front,
the radio-brightness distribution of the SNR would also feature a shell-like morphology.

Although the Vela SNR is in the adiabatic stage of evolution it does not show the usual
behavior of, so called, Sedov SNRs\footnote{The evolution of the typical SNR at the adiabatic
stage in the homogeneous ISM can be described by the Sedov solution \citep{sedov} for a point
explosion.} as described above. Its dynamics is mainly determined by the interaction of the SN
ejecta with numerous clouds with a volume averaged density considerably higher than the density
of the intercloud ISM (\citet{vela_sushch} and references therein). While in the adiabatic Sedov
SNR the hot postshock plasma is a swept up and heated ISM gas, in the Vela SNR the hot postshock
plasma is predominantly the heated and evaporated cloud gas. This difference has a prominent role
in the postshock distribution of relativistic electrons. Due to the low ISM density the transfer
of the SN explosion energy into the ISM (and, in turn, into the cosmic ray acceleration), is small
while the main channel of the ejecta kinetic energy dissipation is the interaction with clouds through
the heating and evaporating of the cloud gas.

Primarily, the dominant process which takes place in the Vela SNR is the interaction between
the ejecta and clouds with generation of the transmitted shock in the cloud and the reflected shock
in the ejecta. Due to the large cloud/intercloud ISM density contrast for the expected Vela SN ejecta
mass of $M_{\mathrm{ej}}\sim 10\,M_{\odot}$ \citep{limongi&chieffi2006} and velocity of 
$V_{\mathrm{ej}} = \sqrt{2E_{SN}/M_{\mathrm{ej}}}\sim 1.5\times 10^3$ km/s the main dissipation 
of the kinetic energy of the ejecta
takes place at reverse shocks generated in the ejecta-clouds interaction.
Due to the low temperature of the ejecta plasma (zero pressure in the analytical treatment
of \citet{truelove_mckee_1999} and $T_{\mathrm{ej}}\sim 10^4$ K in numerical simulations of
\citet{moriya_2013}), the sonic Mach number should be high (up to 100-150 for
$T_{\mathrm{ej}}\sim 10^4$ K) and, thus, the reverse shock will be a region of
effective CR (including relativistic electrons) acceleration. It is more complicated
to estimate parameters of the transmitted shocks, where the transmitted shock velocity
depends on the ejecta and cloud densities $V_{\mathrm{tr}}\sim V_{\mathrm{ej}}
\sqrt{\rho_{\mathrm{ej}}/\rho_{\mathrm{cl}}}$. For the expected cloud radius
$r_{\mathrm{cl}}\sim 0.05$ pc and the number densities $n_{\mathrm{cl}}^{\mathrm{core}}\sim 100\,
\mathrm{cm}^{-3}$ and $n_{\mathrm{cl}}^{\mathrm{corona}}\sim 10\,\mathrm{cm}^{-3}$ in
the two-component approximation in which the cloud consists of a dense core
and a corona around the core \citep{miceli_2006}, at the initial phase of the ejecta-cloud
interaction the velocity of the transmitted shock should be high enough for the
effective particle acceleration, but it will decrease with the distance from the point 
of the SN explosion due to the decrease of the ejecta density.

With time, the main shock will form. The downstream gas will be dominated by the
evaporated plasma with a contribution of the ejecta plasma reheated by the reverse shock. The
contribution of the shocked intercloud plasma is negliglible. Hot postshock plasma will additionally
heat and destroy clouds by thermal conductivity and generating transmitting shocks.

The  total mass of the ejecta and evaporated clouds inside the Vela SNR is about $30\/M_{\odot}$ 
(see Table \ref{phys_params}), i.e. the mass of evaporated clouds is only about twice the 
ejecta mass $M_{\mathrm{ej}}\sim 10\,M_{\odot}$, suggesting that the direct ejecta-cloud
interaction was effective and, in turn, that effective acceleration of relativistic
electrons at strong reverse shocks took place. The close to uniform distribution of clouds
in the ISM leads to a uniform system of strong reverse shocks and, thus, to a nearly uniform
distribution of relativistic electrons inside the Vela SNR.
The close to uniform distribution of the relativistic electrons inside the Vela SNR remains with time due to the
strong turbulent magnetic field (see Section \ref{radio}), which 
restrains the diffusion from the acceleration region. At the same time, the large magnetic field of the Vela SNR (about $50 \mu$G)
does not modify the energy spectrum of relativistic electrons radiated in the range of 30-2700 MHz. The characteristic frequency
of a photon emitted by an electron with energy $\epsilon_{\mathrm{e}}$ is given by \citep{rybicki&lightman}
\begin{equation}
\nu_{\mathrm{ch}} = 0.29 \frac{3q\sin{\alpha}}{4\pi m^{3}c^{5}} \epsilon_{\mathrm{e}}^{2} B \simeq 190 \left[\frac{\epsilon_{\mathrm{e}}}{1\,\mathrm{GeV}}\right]^2 \left[\frac{B}{50\,\mu\mathrm{G}}\right] \mathrm{[MHz]},
\end{equation}
i.e. to emit a 2700 MHz photon an electron with energy $\epsilon_{\mathrm{e},2700}\sim 3.8$ GeV is needed. The cooling time for synchrotron radiation of such electron is 
\citep{b&g}
\begin{equation}
t_{\mathrm{syn}} = 1.3\times10^{6} \left[\frac{\epsilon_{\mathrm{e}}}{4\,\mathrm{GeV}}\right]^{-1} \left[\frac{B}{50\,\mu\mathrm{G}}\right]^{-2} \mathrm{[y]}.
\end{equation}
This time is much longer than the \Ref{estimate of the} age of the Vela SNR \Ref{(regardless the uncertainty which occurs due 
to the low braking index of the pulsar)}, suggesting that electrons can indeed
survive over the time inside the remnant emitting synchrotron radiation.

\subsection{Local discrepancies of the modelled and observed morphology}

The modelled brightness temperature map of the Vela SNR predicts two local elongated peaks, one in the NE part
of the remnant and one in the SW part of the remnant. Observations show that there are two localized peaks in
each part of the remnant. But the identical brightness temperatures of the two NE peaks, Vela\,Y and Vela\,Z, 
and the two SW
peaks, Vela\,W, suggest that physically the two observed peaks in each part of the remnant 
have the same nature and are
two parts of the same peak which could be splitted due to some deviations from our idealised symmetric model.
Another small discrepancy between the modelled
and observed morphologies is that the peaks of Vela\,W are slightly offset from the modelled peak in the SW part of the
remnant. One of the most natural reasons of these discrepancies is that the initial distribution of clouds and,
thus, relativistic electrons does not necessarily have to be uniform.

Both discrepancies can be naturally explained also by the existence of the PWN Vela\,X inside the remnant.
The peak of the radio emission from Vela\,X, as reported by \citet{alvarez2001}, is indicated with a filled square 
in all maps in Fig. \ref{bright_temp}. Indeed, the expansion of the PWN can change the distribution 
of the internal gas and the cloud matter inside the Vela SNR "pushing" them to the outer regions of the remnant. 
This, in turn, would change the distribution of relativistic electrons responsible for the synchrotron radiation,
if they are accelerated at local shocks generated in clouds.
However, the evolution and expansion of Vela\,X is not yet well understood. The PWN features different 
morphologies at different wavebands. At radio and GeV energies an extended ($2\deg \times 3\deg$) "halo" 
emission is observed featuring a "two-wing" structure \citep{velaX_fermi}, which is located mostly to the 
south of the Vela pulsar (as seen in the equatorial coordinates). While the X-ray observations by \emph{ROSAT} 
revealed a much smaller \Refnew{($45^\prime \times 12^\prime$)} emission region ("cocoon") \citep{velaX_rosat}. Subsequent high 
resolution X-ray observations with \emph{Chandra} revealed a structure of the X-ray emission as a composition of 
two toroidal arcs ($17^{\prime \prime}$ and $30^{\prime \prime}$ away from the pulsar) and a $4^\prime$-long collimated jet \citep{velaX_chandra}. 
Finally, in the TeV range, emission spatially coincident with both, halo and cocoon, is detected \citep{velaX_hess_first, hess_velaX}. 
According to the morphology of the halo emission from Vela\,X, the interaction of the PWN with the internal medium 
of the remnant is expected to provide a more important effect on the SW hemisphere of the remnant, since the main part 
of the PWN is located there. Expanding towards the position of Vela\,W peaks, the PWN may cause an increase of the electron 
density and, thus, the enhancement of the synchrotron emission in that region.

Another effect, which may be responsible for the distribution of electrons inside the remnant is the propagation of the reverse
shock inside the remnant. It is argued in the literature that the reverse shock of the SNR may be the reason of the asymmetrical
structure of Vela\,X with respect to the pulsar position \citep{blondin2001}. Due to the difference of the properties of the ambient
medium on the NE and SW sides of the remnant it is possible that the reversed shock was formed earlier in the NE part of the SNR and reached
the PWN Vela\,X suppressing it, while in the SW part of the remnant still no interaction of the PWN with the reverse shock is established
\citep{blondin2001}. If this is the case, then the reverse shock should also influence the distribution of electrons and the intensity
of the emission.



\section{Summary}
\label{sum}

The radio emission from the Vela SNR was simulated in the framework of the hydrodynamical model presented in
\citet{vela_sushch}. This model is based on two hypotheses:
\begin{itemize}
\item the progenitor of the Vela SNR exploded at the border of the stellar wind bubble of the nearby binary system \gammavel\ 
\Refnew{which} causes
the remnant to expand into two media with different densities,
\item the remnant expands into the inhomogeneous medium in which the main bulk of mass is concentrated in small clouds.
\end{itemize}
Originally the model was elaborated to explain a peculiar X-ray emission from the source. In this paper, we showed that the
observed radio flux from the remnant can also be well explained within this model giving it a further observational support.

It was shown that the complicated radio morphology which features several localized emission regions can be explained by the relative
positioning of the Vela SNR and \gammavel, assuming that relativistic electrons responsible for the synchrotron radio emission are
distributed uniformly inside the remnant. The expected observed image of the
Vela SNR depends on how these two objects are positioned relative to each other. We show that for rotation angles $\phi = 35\deg$ and
$\theta = 21\deg$ the expected brightness temperature map of the remnant would feature two peaks in the NE and SW parts of the remnant, which
are coincident with the observed localised emission regions Vela\,Y, Vela\,Z and Vela\,W. The simulated brightness temperatures of the peaks are in
good agreement with the observed brightness temperatures in local emission regions.

We also argue that
the PWN Vela\,X located inside the remnant, which was not taken into account in this study,
may play a notable role in the distribution of relativistic electrons
within the remnant and, thus, in the morphology of the radio emission of the Vela\,Y,
Vela\,Z and Vela\,W regions. The detailed model of the Vela\,X contribution \Ref{to} the
radio emission of Vela SNR will be considered elsewhere.



\begin{acknowledgements}
We would like to thank the referee, Richard Strom, for many useful comments and suggestions, which appreciably improved the paper. 
\end{acknowledgements}

\bibliographystyle{aa}
\bibliography{references}

\end{document}